\documentclass{elsart}

\newtheorem{statement}{Statement}

\begin{document}
\begin{frontmatter}
\title{Supersymmetry a consequence of smoothness?}

\author{H. B. Nielsen}
\address{Niels Bohr Institute, Blegdamsvej 17, DK-2100 Copenhagen \O,
Denmark}
\author{S. Pallua and P. Prester}
\address{Department of Theoretical Physics, PMF, University of
Zagreb, Bijeni\v{c}ka c. 32, 10000 Zagreb, Croatia}
\vspace{0.6cm}
\center{\small{Revised version (first version November 15, 2000)}}

\begin{abstract}
The consequences of certain simple assumptions like smoothness of
gro\-und state properties and vanishing of the vacuum energy (at
least perturbatively) are explored. It would be interesting
from the point of view of building realistic theories to obtain these
properties without supersymmetry. Here we show, however, at least in
some quantum mechanical models, that these simple assumptions lead to
supersymmetric theories.
\end{abstract}
\end{frontmatter}

\section{Introduction}

One may wonder why it is so that the energy spectrum of nature --
locally, i.e. ignoring gravity -- seems to have a bottom, but no top.
Having in mind that there are many parameters -- coupling constants --
which are so far not understood in the sense that we do not have any
theory telling why they should just be what they are, one may ask: If
we varied these parameters/couplings , would the bottom perhaps
disappear? Would the energy density of the ground state -- essentially
the cosmological constant -- remain small?

A major concern of the present article is to claim that assuming the
ground state energy to remain zero -- corresponding to having for all
values zero cosmological constant -- under especially the variation of
the Planck constant $\hbar$ leads in the direction of supersymmetry.

It is of course well known that SUSY theories give zero energy for
the ground state and have been therefore considered as the possible
key to the solution of the small cosmological constant problem (see
\cite{cosmco} for a recent review). SUSY was also shown to have very
simple smoothness properties (see e.g. \cite{Witt81,Witt82,Wittjdg}).
However it is not obvious that there are no non-supersymmetric
field theories with such properties. In fact, that would be even
desirable from the point of view of building realistic models.
Recently there was such an attempt. More precisely, a
nonsupersymmetric string theory was presented which was argued to
have vanishing cosmological constant \cite{KaKuSi} (see also
\cite{Harv98,CoIgKaLa01}). However, the claims in \cite{KaKuSi} were 
criticised by Iengo and Zhu \cite{IeZhu}.

In order to understand if non-SUSY theories with such properties
exist, here we propose to investigate the opposite. We want to start
from some simple assumptions, like vanishing of vacuum energy and/or
certain smoothness properties of ground state, and to consider which
interactions are allowed with these requirements. We shall see that
such assumptions will (at least in cases considered in this paper)
lead us from bosonic theories to SUSY theories with fermion degrees
of freedom.

In fact, we shall start from purely bosonic theories and assume that
properties of the ground state such as energy and wave function are
smooth in parameters of the theory and in particular in Planck
constant $\hbar$. We shall also assume as starting point that the
energy of the ground state vanishes, again at least perturbatively.
This second requirement is in fact not independent from the first
one, because vacuum fluctuations bring kink type singularities and
thus non smooth contributions in $\hbar$.

In our argumentation in the first sections one is
\emph{forced} to put in the fermion degrees of freedom in order to
uphold the (very strong) analyticity/smoothness requirement under
$\hbar$ going even negative. In the case of compact configuration
spaces treated in Section 4 we shall make a very strong 
assumption (but still something reasonable nice to wish for in a
theory good to work with) saying that the classical limit shall work
(perturbatively)
even when formally continuing to negative $\hbar$. Since wave packets
representing classical states tend to jump around under
$\hbar$-changing-sign continuation, we are suggested to identify in the
classical interpretation the startpoint and the endpoint for such
jumps. Thereby strong classical symmetry between different points in
configuration space is to be imposed to uphold the good classical
limit and that is how both an effective fermionic degree of freedom
and SUSY comes in, unavoidably.

\section{Basic assumptions}

Let us start with quantum mechanics of $N$ degrees of freedom,
eventually on the curved space. Then the Schr\"{o}dinger equation
for the ground state wave function reads
\begin{displaymath}
\left(-\frac{\hbar^2}{2}\Delta+V(q)\right)\psi_g(q)=E_g\psi_g(q)
\end{displaymath}
One can rewrite this equation as an equation for $W(q)$, where
\begin{displaymath}
\psi_g(q)=\exp\left(-\frac{W(q)}{\hbar}\right)
\end{displaymath}
and obtain the Riccati equation
\begin{equation}\label{pricc}
V-E_g=\frac{1}{2}\nabla^\mu W\nabla_\mu W-\frac{\hbar}{2}\Delta W
\end{equation}
Our assumptions mean that
\begin{eqnarray}
W(q)&=&\sum_{n=0}^\infty\hbar^nW^{(n)}(q) \label{pwhb} \\
V(q)&=&\sum_{n=0}^\infty\hbar^nV^{(n)}(q) \\
E_g&=&\sum_{n=0}^\infty\hbar^nE_g^{(n)} \label{pehb}
\end{eqnarray}
and in stronger version also $E_g=0$, at least perturbatively. Thus
\begin{displaymath}
V=\frac{1}{2}\nabla^\mu W\nabla_\mu W-\frac{\hbar}{2}\Delta W
\end{displaymath}

For simplicity we shall first consider the case of simple quantum
mechanics on $\mathbf{R}$
\begin{equation}\label{phbos}
H=\frac{p^2}{2}+\frac{1}{2}\left(W'(q)\right)^2-\frac{\hbar}{2}W''(q)
\end{equation}
This is already the bosonic ``half of'' a SUSY
hamiltonian\footnote{For a review of SUSY quantum mechanics see
\cite{CoKhSu}.}. However this
hamiltonian does not satisfy the stability and smoothness assumptions.
In fact, changing continously parameters one can change $W$ in such a
way that makes $\psi_g$ nonnormalizable. An example is $W\sim
ax^{2n}$. If we have finite norm for $a$, that will not be the case
for $-a$. Thus there is no bosonic Hamiltonian in one
dimension satisfying above assumptions. Here a natural generalisation
would be to postulate doubling of the Hilbert space dimension
\begin{displaymath}
\psi(q)\to\left(\begin{array}{c} \psi_1(q) \\ \psi_2(q) \end{array}
\right)
\end{displaymath}
with the new Hamiltonian
\begin{equation}\label{phsusy}
H=\frac{p^2}{2}+\frac{1}{2}\left(W'(q)\right)^2-\frac{\hbar}{2}
W''(q)\sigma_z
\end{equation}
Corresponding to the zero energy eigenstate the wave function may be
written
\begin{equation}\label{pgrsusy}
\psi_g(q)=\exp\left(-\frac{W(q)}{\hbar}\sigma_z\right)\psi(0)
\end{equation}
where $\psi(0)$ is any constant two component column matrix.
Suppose that for some $W$ we have the $\psi_g$ normalizable with
$\sigma_z=+1$, we shall have for $-W$ normalizable state with
$\sigma_z=-1$. The above Hamiltonian is the well known SUSY Hamiltonian
where function $W$ is called superpotential
\cite{Witt81,ClaHal,CoKhSu}.

In the compact case ($S^1$ configuration space) we cannot use the
normalization requirement. However in this case $W$ has at least two
critical points and the Hamiltonian (\ref{phbos}) has two independent
degenerate (normalizable) ground state wave functions
$\psi_g(q)\propto\exp(\mp W(q)/\hbar)$.
Both critical points are classical minima where $W'(q)=0$.
However, one is maximum of $W$ and another minimum of $W$. Due to
this the ratio of probabilities to find particle in classical minima
is
\begin{displaymath}
\left(\frac{\psi_g(q_1)}{\psi_g(q_2)}\right)^2=\exp\left\{
\frac{2}{\hbar}\left[W(q_2)-W(q_1)\right]\right\}
\end{displaymath}
This particle is concentrated around minimum of $W$. Changing sign
of $\hbar$ the situation will reverse. Again this smoothness problem
will be avoided if we add an internal degree of freedom and write
Hamiltonian as in (\ref{phsusy}). The ground state wave function will
now again be (\ref{pgrsusy}). Now the change of sign of $\hbar$ will
change only the sign of $\sigma_z$ eigenvalue, but the probabilities
of position in configuration space will stay the same.

The smoothness assumption can also lead to $E_g=0$, at least to
first order in $\hbar$. This is due to the fact that the usual
quantum fluctuations violate smoothness property. For instance, for
the harmonic oscillator $E_q=|\hbar\omega|/2$.
One would need a compensating term, for instance $-\hbar\omega/2$.
This however would be satisfactory for one sign of $\hbar$ only.
What would help is to add a term $\hbar\omega\sigma_z/2$ which
depends on extra degree of freedom $\sigma_z$ that can take two
values $\sigma_z=\pm1$
\begin{displaymath}
H=\frac{p^2}{2}+\frac{1}{2}\omega^2q^2-\frac{1}{2}\hbar\omega\sigma_z
\end{displaymath}
The ground state energy $E_g$ is now vanishing both for positive and
negative $\hbar$.

\section{SUSY from perturbative vanishing of cosmological constant}


It is well known that one of the consequences of non-renormalization
theorems in SUSY QFT is that if effective potential vanishes at some
point at the tree level, then it vanishes at that point to all
finite orders of perturbation theory \cite{Stevie3}. That means that
in SUSY theories the zero energy property of the classical vacuum is
not changed by higher orders in perturbation theory.

Now, let us reverse above statement in the following way. Suppose
that we start with a system which has only bosonic degrees of freedom
$\varphi_i$ and some generic regular potential $V(\varphi)$ (unless
stated differently we assume bounding potential, i.e.,
$V(\varphi)\to\infty$ as $|\varphi_i|\to\infty$). Classical
(tree level) vacua are minima of $V(\varphi)$. Let us now insist that
for some reason (e.g., small cosmological constant) classical vacua
are isolated and have
energy equal to zero \emph{genericaly}, i.e., for general values of
parameters appearing in $V(\varphi)$. Now, $V$ is positive so it can
be written as $V(\varphi)=\sum_n(f_n(\varphi))^2$,
where $f_n$ are some generic regular functions. To obtain only a
discrete set of vacua, the number of functions $f_n$ should be equal to
the dimension of the configuration space.

We shall take $f_n(\varphi)$ equal to a gradient of some function
$W(\varphi)$. In fact this follows from the Riccati equation
(\ref{pricc}) and smoothness assumptions (\ref{pwhb}--\ref{pehb}).

After quantization, quantum fluctuations will move energies of the
vacua to non-zero values. Now we come to the central question. Could we
obtain SUSY (or its part) just adding something in a ``minimal'' way
to keep all classical vacua perturbatively at zero energy?
By ``minimaly'' we think adding minimal number of new degrees of
freedom and preserving asymptotic behaviour of bosonic potential.
We will analyse this through a few simple examples.

\subsection{One-dimensional QM}

Let us first consider the simplest model, a particle moving in 1-D
space under influence of the external potential. The Hamiltonian is
given with
\begin{displaymath}
H=\frac{1}{2}p^2+\frac{1}{2}(W'(x))^2
\end{displaymath}
where the mass is scaled to one and prime denotes derivation.
The ``superpotential'' $W$ will genericaly have a number of
nondegenerate critical points $x^{(i)}$, any of which correspond to
classical zero-energy state. 

Consider perturbation theory around one classical minimum $x^{(i)}$.
In the lowest order (harmonic oscillator) approximation we have
\begin{displaymath}
H_0=\frac{1}{2}p^2+\frac{1}{2}\left(W''(x^{(i)})\right)^2
\left(x-x^{(i)}\right)^2
\end{displaymath}
We can see that lowest order quantum correction of classical
zero-energy state is
\begin{displaymath}
E_0^{(i)}=\frac{\hbar}{2}|W''(x^{(i)})|
\end{displaymath}
so the corrected Hamiltonian $H_0^c$, with zero-energy ground state
in harmonic oscillator approximation, is
\begin{equation} \label{hc0}
H_0^c=\frac{1}{2}p^2+\frac{1}{2}\left(W''(x^{(i)})\right)^2
\left(x-x^{(i)}\right)^2-\frac{\hbar}{2}\left|W''(x^{(i)})\right|
\end{equation}
Now comes the central question: can we generalize the above
construction, i.e., find corrected Hamiltonian $H^c$ for which
classical ground states continue to have zero energy to any finite
order of perturbation theory? Naively, a simple generalisation of Eq.
(\ref{hc0}) could be
\begin{equation}\label{hcsqrt}
H^c=\frac{1}{2}p^2+\frac{1}{2}\left(W'(x)\right)^2-
\frac{\hbar}{2}\left|W''(x)\right|
\end{equation}
or
\begin{equation}\label{hcsign}
H^c=\frac{1}{2}p^2+\frac{1}{2}\left(W'(x)\right)^2-
\frac{\hbar}{2}\,\mbox{sign}\!\left(W''(x^{(i)})\right)W''(x)
\end{equation}
Now we argue that both suggestions are unacceptable:
\begin{itemize}
\item (\ref{hcsqrt}) and (\ref{hcsign}) both contain a square root
which is ``non-analytical''
\item (\ref{hcsqrt}) doesn't do the trick at next order of
perturbation theory. Indeed, it is shown in Appendix A that
(\ref{hcsign}) is a unique form if we allow that $W$ depends on $\hbar$
(i.e. other solutions can, after redefining $W$, be put in the form
(\ref{hcsign})).
\item if we order critical points such that $x^{(i)}<x^{(i+1)}$
Hessians $W''(x^{(i)})$ will have alternating signs, so correction
term in (\ref{hcsign}) will do the trick only in every second
critical point. We loose half of classical zero-energy vacua.
\end{itemize}
The ``minimal'' way to keep all wanted properties, i.e.,
perturbative preserving of zero energy for all classical vacua and
keeping asymptotic behaviour of potential, is to take
\begin{equation}\label{hc}
H^c=\frac{1}{2}p^2+\frac{1}{2}\left(W'(x)\right)^2+
\frac{\hbar}{2}W''(x)\sigma
\end{equation}
where $\sigma$ is an operator which acts on some internal space of
states (commutes with $x$ and $p$) and have one +1 and one -1
eigenvalue, so that there is state
with correct sign for every critical point. Obviously, a ``minimal''
choise is to take two-dimensional internal space and one of Pauli
matrices for $\sigma$, e.g., $\sigma^z$. The thus obtained Hamiltonian
(\ref{hc}) is just Witten's $N=2$ SUSY QM.

\subsection{$D$-dimensional QM}

Consider now $D$-dimensional version of above model. We start from
the bos\-on\-ic Hamiltonian
\begin{equation}\label{hD}
H=\frac{1}{2}\sum_{a=1}^{D}\left[(p_a)^2+\left(\partial_a
W(x)\right)^2\right] \qquad a=1,\ldots,D
\end{equation}
and try to apply the same argumentation as in previous subsection.

The ``Superpotential'' $W(x)$ generaly has a number of nondegenerate
critical points. We expand the potential around one of them and in
the lowest order get
\begin{displaymath}
H_0=\frac{1}{2}p_ap_a+\frac{1}{2}\partial_a\partial_bW(x^{(i)})
\partial_a\partial_cW(x^{(i)})
\left(x-x^{(i)}\right)_b\left(x-x^{(i)}\right)_c
\end{displaymath}
Now we make a rotation\footnote{In this and following subsection
primes denote coordinate systems. In the rest of the paper prime
denotes derivation of the function.} $x'=\mathcal{O}^{(i)}x$,
$\mathcal{O}^{(i)}\in O(D)$ that diagonalizes the Hessian of $W(x)$
at $x^{(i)}$. In the rotated coordinates $H_0$ is (we denote $W'(x')
\equiv W(x)$)
\begin{displaymath}
H_0=\frac{1}{2}\sum_{a=1}^{D}\left[(p'_a)^2+\left(\partial'^2_a
W'(x'^{(i)})\right)^2\left(x'_a-x'^{(i)}_a\right)^2\right]
\end{displaymath}
These are just $D$ decoupled harmonic oscillators, so the ground
state energy is 
\begin{displaymath}
E_0^{(i)}=\frac{\hbar}{2}\sum_{a=1}^{D}\left|\partial'^2_a
W'(x'^{(i)})\right|
\end{displaymath}
Again, by the simmilar argument as in the one-dimensional
case, we can conclude there is no $D$-dimensional bosonic Hamiltonian
satisfying our assumptions.

As in the previous section we can construct in the first order
approximation the corrected Hamiltonian which preserves zero-energy
classical ground state connected to $i$-th critical point. It is
given with:
\begin{equation}\label{Dhc0}
H^c_0=\frac{1}{2}\sum_{a=1}^{D}\left[(p'_a)^2+\left(\partial'^2_a
W'(x'^{(i)})\right)^2\left(x'_a-x'^{(i)}_a\right)^2+\hbar
\partial'^2_aW'(x'^{(i)})\sigma^{(i)}_a\right]
\end{equation}
where $\sigma^{(i)}_a$ are hermitian operators on some internal
space which have following properties:
\begin{enumerate}
\item $\left(\sigma^{(i)}_a\right)^2=1$
\item $\mbox{Tr}\,\sigma^{(i)}_a=0$
\item $[\sigma^{(i)}_a,\sigma^{(i)}_b]=0$ for $a\ne b$
\end{enumerate}
which follow from the ``minimal'' requirement that $\sigma^{(i)}_a$
should act only as signs so that we obtain one and only one
zero-energy vacuum near any critical point of $W$ (regardless of
its index). It is clear that the smallest representation of the
above algebra is defined on $2^D$ dimensional state composed od $D$
independent ``sites'' on which $\sigma^{(i)}_a$ act as Pauli
matrices.

Now, for the full corrected Hamiltonian we could naively
expect (still in rotated coordinates $x'$):
\begin{displaymath}
H^c=\frac{1}{2}\sum_{a=1}^{D}\left[(p'_a)^2+\left(\partial'_a
W'(x')\right)^2+\hbar\partial'^2_aW'(x')\sigma^{(i)}_a\right]
\end{displaymath}
But, we can immidiately see a problem; the rightmost term
(``correction'') is not invariant under rotations, so in a generic
coordinate system it would have non-diagonal terms ($\partial_a
\partial_b W$). Because non-diagonal terms obviously do not enter
in first order $H^c_0$ (see (\ref{Dhc0})), it is natural to include
them also. So, we are led to the form:
\begin{equation}\label{Dhci}
H^c=\frac{1}{2}\left[p'_ap'_a+\partial'_aW'(x')\partial'_aW'(x')+
\hbar\partial'_a\partial'_bW'(x')\chi^{(i)}_{ab}\right]
\end{equation}
where $\chi^{(i)}_{ab}$ are operators such that
$\chi^{(i)}_{aa}=\sigma^{(i)}_a$, but still unspecified for $a\ne b$.

Now, we want from $H^c$ to have zero-energy perturbative vacuum at
every critical point, and not just $i$-th. If we make the above
procedure for some other critical point $j\ne i$ we would obtain
\begin{equation}\label{Dhcj}
H^c=\frac{1}{2}\left[p''_ap''_a+\partial''_aW''(x'')\partial''_aW''(x'')+
\hbar\partial''_a\partial''_bW''(x'')\chi^{(j)}_{ab}\right]
\end{equation}
where $x''=\mathcal{O}^{(j)}x$ are coordinates which diagonalizes
the Hessian of $W$ at $x^{(j)}$, $W''(x'')=W(x)$, and
$\chi^{(j)}_{ab}$ are operators with the same properties as
$\chi^{(i)}_{ab}$. Now, (\ref{Dhci}) and (\ref{Dhcj}) should be two
forms of the \emph{same} Hamiltonian, written in different coordinate
systems connected with rotation
\begin{displaymath}
x''=\mathcal{O}^{(ij)}x',\qquad
\mathcal{O}^{(ij)}=\mathcal{O}^{(j)}\left(\mathcal{O}^{(i)}\right)^\top
\end{displaymath}
If we now write Hamiltonian (\ref{Dhcj}) in $x'$ coordinates and
compare it with (\ref{Dhci}) we obtain condition
\begin{displaymath}
\chi^{(j)}_{ab}=\mathcal{O}^{(ij)}_{ac}\mathcal{O}^{(ij)}_{bd}
\chi^{(j)}_{cd}
\end{displaymath}
We want this to be true for any pair of critical points for generic
$W(x)$. That leads us to the following conclusion: if we can find
operators $\chi_{ab}$ such that sets $\chi_{aa}$ and $\chi'_{aa}$
both satisfy conditions 1-3, where primed set is obtained by
applying arbitrary rotation $\mathcal{O}\in O(D)$, i.e.
\begin{equation}\label{Dcond}
\chi'_{ab}=\mathcal{O}_{ac}\mathcal{O}_{bd}\chi_{cd}
\end{equation}
then the Hamiltonian given with
\begin{equation}\label{Dhc}
H^c=\frac{1}{2}\left[p_ap_a+\partial_aW(x)\partial_aW(x)+
\hbar\partial_a\partial_bW(x)\chi_{ab}\right]
\end{equation}
will have required properties.

It is obvious that SUSY case $\chi_{ab}=[\bar{\psi}_a,\psi_b]$
satisfy above condition. Now we shall explicitly show for $D=2$ that
this is a unique solution.

\subsubsection{$D=2$ QM}

We can take the set of Hermitian operators $\chi_{ab}$ to be
symmetric in indeces, so there are three independent operators,
$\chi_{11}$, $\chi_{22}$ and $\chi_{12}$. Without any loss of
generality we can take a following representation for diagonal
operators:
\begin{displaymath}
\chi_{11}=\sigma_1^z=\sigma^z\otimes1, \qquad
\chi_{22}=\sigma_2^z=1\otimes\sigma^z
\end{displaymath}
General matrix $\mathcal{O}\in O(2)$ has standard parametrization:
\begin{displaymath}
\mathcal{O}=\left(\begin{array}{cc} \cos\phi & \sin\phi \\ 
-\sin\phi & \cos\phi \end{array}\right)
\end{displaymath}
Using this in (\ref{Dcond}) gives
$\chi'_{11}=\mathcal{O}_{1c}\mathcal{O}_{1d}\chi_{cd}$,
which inserted in condition $(\chi'_{11})^2=1$ gives
\begin{displaymath}\!\!\!\!\!\!
\sin(2\phi)\!\left(2\chi_{12}^2+\chi_{11}\chi_{22}-1\right)+2\cos^2\!\phi
\,\{\chi_{11},\chi_{12}\}+2\sin^2\!\phi\,\{\chi_{22},\chi_{12}\}=0
\end{displaymath}
If we take $\phi=0,\pi/2$, we obtain conditions:
\begin{displaymath}
0=\{\chi_{11},\chi_{12}\}=\{\chi_{22},\chi_{12}\}=2\chi_{12}^2+
\chi_{11}\chi_{22}-1
\end{displaymath}
Explicit calculation in the above representation gives
\begin{displaymath}
\chi_{12}=\left(\begin{array}{cccc} 0 & 0 & 0 & 0 \\ 0 & 0 & \omega
& 0 \\ 0 & \omega^* & 0 & 0 \\ 0 & 0 & 0 & 0 \end{array}\right),
\qquad |\omega|=1
\end{displaymath}
Now we'll show that we have in fact obtained $N=2$ SUSY QM. To do
so, we first introduce fermionic operators $\psi_a$ and
$\bar{\psi}_a=\psi_a^\dagger\,$, $a=1,2$, which satisfy CAR
$\{\bar{\psi}_a,\psi_b\}=\delta_{ab}$. We can represent them using
Jordan-Wigner transformation:
\begin{displaymath}
\psi_1=\sigma_1^-=\sigma^-\otimes1,\qquad
\psi_2=-\omega\sigma_1^z\sigma_2^-=-\omega\sigma^z\otimes\sigma^-
\end{displaymath}
From that follows
\begin{displaymath}
[\bar{\psi_1},\psi_2]+[\bar{\psi_2},\psi_1]=2\omega\sigma_1^+
\sigma_2^-+2\omega^*\sigma_1^-\sigma_2^+=2\left(\begin{array}{cccc}
0 & 0 & 0 & 0 \\ 0 & 0 & \omega & 0 \\ 0 & \omega^* & 0 & 0 \\
0 & 0 & 0 & 0 \end{array}\right)=2\chi_{12}
\end{displaymath}
which shows that we can write corrected Hamiltonian (\ref{Dhc}) in
the form
\begin{displaymath}
H^c=\frac{1}{2}\left(p_ap_a+\partial_aW(x)\partial_aW(x)+
\hbar\partial_a\partial_bW(x)[\bar{\psi_a},\psi_b]\right)
\end{displaymath}
i.e., $N=2$ SUSY QM \cite{ClaHal}.

\subsection{Wess-Zumino QM}

Consider now QM model on the complex plane, with Hamiltonian given
with:
\begin{equation}\label{wzh}
H=\bar{p}p+\bar{\partial}\bar{W}(\bar{z})\partial W(z)
\end{equation}
where $W(z)$ is holomorphic function, and CCR are $[z,p]=i\hbar$,
$[\bar{z},\bar{p}]=i\hbar$. Obviously, if we write $z=(x_1+ix_2)/
\sqrt{2}$, $\partial=(\partial_1-i\partial_2)/\sqrt{2}$,
and use Cauchy-Riemann conditions, we can see that (\ref{wzh}) is a
special case of $D=2$ Hamiltonian (\ref{hD}) with
$u(x)=2\mbox{Re}\,W(z)$ playing the role of $W(x)$ in (\ref{hD}).
The only difference from the analysis in last subsection comes
from the fact that $u(x)$ satisfies Laplace equation (consequence of
Cauchy-Riemann conditions):
\begin{displaymath}
\sum_{a=1}^{2}\partial_a^2 u(x)=0
\end{displaymath}
It means that all critical points of $u(x)$ have index -1 and
traceless Hessians, so we must take this as a ``constraint'' (in the
last subsection we supposed generic $W(x)$ with generic critical
points. From this follows that instead of (\ref{Dhc0}) we now have
around $i$-th critical point
\begin{displaymath} \!\!\!\!\!\!\!\!\!
H^c_0=\frac{1}{2}\sum_{a=1}^{D}\left[(p'_a)^2+\left(\partial'^2_a
u'(x'^{(i)})\right)^2\!\left(x'_a-x'^{(i)}_a\right)^2\right]+
\frac{\hbar}{2}\left(\partial'^2_1u'-\partial'^2_2
u'\right)\!(x'^{(i)})\,\sigma^{(i)}
\end{displaymath}
where again $\left(\sigma^{(i)}\right)^2=1$. We can again conclude
that corrected Hamiltonian is
\begin{equation}\label{wzhc}
H^c=\frac{1}{2}\left[p_ap_a+\partial_au(x)\partial_au(x)+
\hbar\partial_a\partial_bu(x)\chi^{(i)}_{ab}\right]
\end{equation}
but where operators $\chi_{ab}$ now satisfy the properties
$\chi_{11}=-\chi_{22}\equiv\sigma$, $\sigma^2=1$, which are
invariant on $O(2)$ rotations (i.e., $\chi'_{ab}=\mathcal{O}_{ac}
\mathcal{O}_{bd}\chi_{cd}$ should have the same properties for any
$\mathcal{O}\in O(2)$).

Now, from $\left(\chi'_{11}\right)^2=1$ we obtain condition
\begin{displaymath}
\sin(2\phi)\left(\chi_{12}^2-1\right)+\cos(2\phi)\left\{\sigma,
\chi_{12}\right\}=0
\end{displaymath}
Taking $\phi=0,\pi/4$ it follows that $\chi_{12}$ should satisfy
$\{\sigma,\chi_{12}\}=0$, $(\chi_{12})^2=1$.
We can represent obtained algebra with
\begin{equation}\label{wzchi}
\sigma=\chi_{11}=-\chi_{22}=\sigma^x\,,\qquad \chi_{12}=\sigma^y
\end{equation}
Obtained model is \emph{not} SUSY, but we shall now show how it is
connected to SUSY.

It is well-known \cite{ClaHal,Mathur,JaLeLe} that there is $N=4$ SUSY
model connected to (\ref{wzh}):
\begin{equation}\label{wzqm}
H_{WZ}=\bar{p}p+\bar{\partial}\bar{W}(\bar{z})\partial W(z)+\hbar
\partial^2W(z)\psi_2\psi_1+\hbar\bar{\partial}^2\bar{W}(\bar{z})
\bar{\psi}_1\bar{\psi}_2
\end{equation}
which is also a special case of $N=2$ SUSY QM in $D=2$ (\ref{Dhc}).
It is named Wess-Zumino QM because it can be obtained by dimensional
reduction from Wess-Zumino SUSY QFT in two dimensions (every SUSY QM
model obtained from SUSY QFT by dimensional reduction has at least
$N=4$ SUSY). Let's consider properties of WZ QM. First, we define a
vacuum $|+\rangle$ and a fully filled fermion state $|-\rangle$ as
\begin{displaymath}
\psi_a|+\rangle=0\,,\qquad |-\rangle\equiv\bar{\psi}_1\bar{\psi}_2
|+\rangle
\end{displaymath}
These states span bosonic part of the Hilbert space (convention).
It is easy to se that $H_{WZ}$ acts on fermionic part of Hilbert
state (spaned by $\bar{\psi}_a|+\rangle$) as (\ref{wzh}) which is
positive definite so there are no zero energy fermionic states. It
appears that when SUSY cannot make zero energy states in some
sector, then it does \emph{nothing} in that sector.

Let us now analyze action of $H_{WZ}$ on bosonic states. It is easy
to see that if we use a representation
\begin{displaymath}
|+\rangle={1 \choose 0}\,\qquad |-\rangle={0 \choose 1}
\end{displaymath}
then we have $\psi_2\psi_1=\sigma^+$ and
$\bar{\psi}_1\bar{\psi}_2=\sigma^-$,
so the ``SUSY part'' of $H_{WZ}$ is represented with
\begin{displaymath}
\partial^2W(z)\psi_2\psi_1+\bar{\partial}^2\bar{W}(\bar{z})
\bar{\psi}_1\bar{\psi}_2=\partial_1^2u(x)\sigma^x+\partial_1\partial_2
u(x)\sigma^y
\end{displaymath}
We now see that $H_{WZ}$ constrained on bosonic part of Hilbert
state (where there are all zero-energy vacua) is equal to our
corrected Hamiltonian (\ref{wzhc}), (\ref{wzchi}).

\section{``Als ob'' fermions from bosons}

\subsection{Quantum mechanics in compact space}

We consider particle moving in $D$-dimensional compact Riemann space
with the Hamiltonian given with
\begin{displaymath}
H=-\hbar^2\nabla^2+V(q)
\end{displaymath}
where $q=(q_1,\ldots,q_D)$ are coordinates, and $\nabla$ the covariant
derivative. Again, as in the Section 2, the potential $V(q)$ can be
written in the form of Riccati equation
\begin{equation} \label{Dricc}
V(q)-E_g=(\nabla W(q))^2-\hbar\nabla^2W(q)
\end{equation}
where $E_g$ is the ground state energy, and $W(q)$ is connected
to the ground state wave function $\psi_g(q)$ with
\begin{displaymath}
W(q)=-\hbar\ln\psi_g(q)
\end{displaymath}
We want again for our system to have ``smooth'' classical limit, so
we take $V(q)$, $E_q$ and $W(q)$ to be ``expandable'' in $\hbar$:
\begin{eqnarray}
V(q)&=&\sum_{n=0}^\infty\hbar^nV^{(n)}(q) \label{vexph} \\
E_g&=&\sum_{n=0}^\infty\hbar^nE_g^{(n)} \nonumber \\
W(q)&=&\sum_{n=0}^\infty\hbar^nW^{(n)}(q) \label{wexph}
\end{eqnarray}
This leads us to two statements.
\begin{statement}
The classical potential $V^{(0)}(q)$ has at least two equally deep
minima, i.e., there exists at least two points $q_i$ for which
$V^{(0)}(q_i)-E_g^{(0)}=0$. More precisely, number of these
classical minima is equal to the number of critical points of
$W^{(0)}(q)$.
\end{statement}
\begin{pf*}{Proof.}
If we put (\ref{vexph}--\ref{wexph}) into
(\ref{Dricc}) and take $\hbar=0$, we obtain
\begin{equation} \label{clricc}
V^{(0)}(q)-E_g^{(0)}=\left(\nabla W^{(0)}(q)\right)^2
\end{equation}
But now, on a compact configuration space every function has at
least one local minimum and one local maximum (and so $W^{(0)}$), so
there are at least two points $q_i$ in which $\nabla W^{(0)}=\partial
W^{(0)}=0$. Inserted in (\ref{clricc}) that proves the statement.
\end{pf*}
\begin{statement}
Main concentration of probability for the ground state (measured by
$|\psi_g(q)|^2$) will jump from around the global maximum to around
the global minimum of $W(q)$ when $\hbar$ is continously passing by
$\hbar=0$. 
\end{statement}
\begin{pf*}{Proof.} It is clear that for $\hbar$ small enough critical
points of $W(q)$ just have slightly different positions from those of
$W^{(0)}(q)$. For small $\hbar$ ground state wave function is
approximately 
\begin{displaymath}
\psi_g(q)=\exp\left[-\frac{1}{\hbar}\sum_{n=0}^\infty\hbar^nW^{(n)}
(q)\right]\approx\exp\left[-\frac{1}{\hbar}W^{(0)}(q)\right]
\end{displaymath}
It is clear that for $\hbar$ positive (negative) $\psi_g(q)$ is
sharply peaked around the global maximum (minimum) of $W^{(0)}(q)$,
so we have a ``jump'' when $\hbar$ is passing zero.
\end{pf*}

It is the crux of our ``derivation'' of (need for) SUSY that we
declare:

\emph{Such a ``jumping'' under $\hbar$ passing $\hbar=0$ (from
$\hbar>0$ to $\hbar<0$) means that the classical limit is not good!}

That is to say, we assume that we shall be able -- if needed using
some slightly modified identification of states with classical states
-- to arrange this ``jumping'' to be avoided. If not, we do not
accept the system as obeying what we loosely call ``our smoothness
assumptions''.

Our ``solution'' to the
jumping-of-states-to-different-minimum-of-$V$-problem is proposed
as:

\emph{We propose to change the classical configuration space by
putting together to one point so many points as are needed to have
all the ``jumps'' for $\hbar\to-\hbar$ occur between original
$q$-points now identified to be interpreted as only one
point.}

If we want to have classical physics not to distinguish the points
to be identified -- say we identify $q\to f(q)$ --
then at least to classical approximation we must have
\begin{enumerate}
\item The map $f:\mbox{configuration space}\to\mbox{configuration
space}$ being an isometry for the metric $g_{ab}(q)$ of the kinetic
term
\begin{displaymath}
g_{ab}(f(q))\frac{\partial f^a}{\partial q^c}\frac{\partial f^b}
{\partial q^d}=g_{cd}(q)
\end{displaymath}

\item $V(f(q))=V(q)$
\end{enumerate}

We expect that additional variables, introduced to denote different
(bosonic) configurations which are classically indistinguishable,
will behave as fermionic degrees of freedom, at least locally around
classical vacua, or in perturbative expansion.

\subsubsection{Example: a circle}

As a simple example of the above ideas, let us consider one
dimensional particle on a flat circle\footnote{In one dimension all
metric tensor fields can be made trivial, i.e. $g_{11}=1$, by
appropriate choice of coordinate.}. The Hamiltonian is now
\begin{equation} \label{hcirc}
H=p^2+V(q)
\end{equation}
where $q\sim q+4\pi$ (we denote the configuration space $S^1_{4\pi}$).
In the simplest case there are two classical vacua. It follows that
there are only two possible isometric maps $f$
\begin{eqnarray}
f(q)&=&q+2\pi\pmod{4\pi} \label{trans} \\ f(q)&=&2\pi-q\pmod{4\pi}
\nonumber
\end{eqnarray}
(this follows from $f(f(q))=q\pmod{4\pi}$). By arguing about a
slightly pushed ground state -- a superposition of the ground state
and first excited state -- we may argue for (\ref{trans}).

If we take for granted that the points on the $S^1_{4\pi}$ to be
identified are
\begin{displaymath}
q\longleftrightarrow f(q)=q+2\pi \pmod{4\pi}
\end{displaymath}
we may look for an operator $Q$ that maps the state $\psi:S^1_{4\pi}
\to\mathbf{C}$ into the another state localized at ``same classical
points'' (but at different $q$, namely $f(q)$). More precisely, we
want that if $\psi$ is a $\hat{q}$-eigenstate, say $\psi(q)=\delta
(q-q_0)$, then $Q\psi$ should be nonzero only at $q_0$ and $f(q_0)$.
The requirement of $Q\delta(q-q_0)$ being local in the sense of
$\delta$-functions and their derivatives around $q_0$ or $q_0+2\pi$
means
\begin{eqnarray*}
Q\delta(q-q_0)&=&P_2(\hat{p},\hat{q})e^{i\frac{2\pi\hat{p}}{\hbar}}
\delta(q-q_0)+P_1(\hat{p},\hat{q})\delta(q-q_0) \\
&=&P_2(\hat{p},\hat{q})``\sigma_x"\delta(q-q_0)+P_1(\hat{p},\hat{q})
\delta(q-q_0)
\end{eqnarray*}
where $``\sigma_x"$ is the translation operator by $2\pi$, i.e.,
\begin{displaymath}
``\sigma_x"\equiv\exp\left(\frac{i}{\hbar}2\pi\hat{p}\right)
\end{displaymath}
and $P_\alpha$, $\alpha=1,2$ are finite polinomials in $\hat{p}$ (so
they can only make infinitesimal translations) and smooth
$4\pi$-periodic functions of $\hat{q}$.

Using the fact that $\hat{q}$-eigenstates $\delta(q-q_0)$ constitute
a complete basis we argue that the operator $Q$ is of the form
\begin{displaymath}
Q=P_2(\hat{p},\hat{q})``\sigma_x"+P_1(\hat{p},\hat{q})
\end{displaymath}
We are tempted to drop the second term because we really want the
part of $Q$ that shift the state from one $q$-neighbourhood to
another one (around $q+2\pi \pmod{4\pi}$).

Let us return to the system which we want to analyze, with
Hamiltonian given in (\ref{hcirc}). Using Riccati eq.
(\ref{Dricc}) it can be written as
\begin{equation} \label{hfac} \!\!\!\!\!\!\!\!
H=\hat{p}^2+\left(W'(\hat{q})\right)^2-\hbar W''(\hat{q})+E_g=\left(
\hat{p}+iW'(\hat{q})\right)\left(\hat{p}-iW'(\hat{q})\right)+E_g
\end{equation}

From the requirement that the classical potential $V^{(0)}(q)$, for
which Riccati equation gives
\begin{displaymath}
V^{(0)}(q)=\left(W^{(0)\,\prime}(q)\right)^2
\end{displaymath}
is the same for $q$ and $f(q)=q+2\pi$, follows
\begin{displaymath}
\left.\left(\frac{dW^{(0)}}{dq}\right)^2\,\right|_q=\left.\left(
\frac{dW^{(0)}}{dq}\right)^2\,\right|_{q+2\pi}
\end{displaymath}
which means that $W^{(0)}$ must be antiperiodic with period $2\pi$.
We now assume the same property for the full $W$, i.e.,
\begin{equation} \label{waper}
W(q)=-W(q+2\pi) \pmod{4\pi}
\end{equation}
This can also be written using $``\sigma_x"$ as
\begin{equation}\label{acsigw}
\left\{``\sigma_x",W(\hat{q})\right\}=0
\end{equation}
From $4\pi$-periodicity follows 
\begin{displaymath}
\left(``\sigma_x"\right)^2=1 \qquad \left(``\sigma_x"\right)^\dagger=
``\sigma_x"
\end{displaymath}
For the completeness we add also trivial relation
\begin{equation}\label{acsigp}
\left[``\sigma_x",\hat{p}\,\right]=0
\end{equation}
Now if we take for $Q$
\begin{equation}\label{qcirc}
Q=Q^\dagger=\left(\hat{p}+iW'(\hat{q})\right)``\sigma_x"
\end{equation}
from (\ref{hfac}) and (\ref{acsigw}--\ref{acsigp}) follows 
\begin{equation}\label{n1susy}
H=\frac{1}{2}\{Q,Q\}+E_g=Q^2+E_g
\end{equation}
If we could find fermion number operator $F$ such that $(-1)^F$
anticommutes with $Q$, we could say that our starting \emph{purely
bosonic} system can be written as supersymmetric. This is our next
task.

Locally in $q$, or perturbatively, we define a fermion number $F$ so
that
\begin{equation}\label{fernum}
(-1)^F=``\sigma_z"
\end{equation}
where $``\sigma_z"$ is defined for the neighbourhoods (of trivial
topology) of critical points of $W(q)$ (which are near classical
vacua for $\hbar$ small) in the following way. Denote by $q_g$
minimum of $W(q)$ and arrange that $0<q_g<2\pi$. Because of the
$2\pi$-antiperiodicity of $W$ we know that maximum of $W$ is at
$f(q_g)=q_g+2\pi$. Now, equivalence $q\sim f(q)$ reduces classical
configuration space from $S^1_{4\pi}$ to $S^1_{2\pi}=S^1_{4\pi}/
\mathbf{Z_2}$. Because quantum corrections break the equivalence,
beside ``classical position'' $\bar{q}\in[0,2\pi]$ we need another
discrete degree of freedom which tells us in which of the classically
equivalent points ($\bar{q}$ or $\bar{q}+2\pi$) particle is. More
formaly, we split the wave function $\psi(q)$, $q\in[0,4\pi\rangle$
in two components $\psi(\bar{q},\sigma)$, $q\in[0,2\pi]$,
$\sigma=\pm1$ in the following way
\begin{equation}\label{splcp}
\psi(\bar{q},1)\equiv\psi(\bar{q}),\quad
\psi(\bar{q},-1)\equiv\psi(\bar{q}+2\pi),\qquad
\bar{q}\in[0,2\pi]
\end{equation}
From the definition of $``\sigma_x"$ follows
\begin{displaymath}
``\sigma_x"\psi(q)=\psi(q+2\pi)
\end{displaymath}
so we have 
\begin{displaymath}
``\sigma_x"\psi(\bar{q},\sigma)=\psi(\bar{q},-\sigma),\qquad
\sigma=\pm1
\end{displaymath}
We can now define operator $``\sigma_z"$ such that
\begin{equation}\label{sigz}
``\sigma_z"\psi(\bar{q},\sigma)=\sigma\psi(\bar{q},\sigma)
\end{equation}
Obvious properties of $``\sigma_z"$ are
\begin{displaymath}
\{``\sigma_z",``\sigma_x"\}=[``\sigma_z",\hat{p}\,]=
[``\sigma_z",\bar{q}\,]=0
\end{displaymath}
\begin{displaymath}
\left(``\sigma_z"\right)^2=1,\qquad \left(``\sigma_z"\right)^\dagger
=``\sigma_z"
\end{displaymath}
From that, (\ref{qcirc}), and (\ref{fernum}) trivially follows
\begin{displaymath}
\{(-1)^F,Q\}=\{``\sigma_z",Q\}=0
\end{displaymath}

Finally, using $\bar{q}$ and $``\sigma_z"$ instead of $q$, we can
formally write Hamiltonian (\ref{hfac}) in the standard $N=2$ SUSY
form
\begin{equation}\label{h2pi}
H=\hat{p}^2+\left(W'(\hat{\bar{q}})\right)^2-\hbar
W''(\hat{\bar{q}})``\sigma_z"+E_g
\end{equation}
where we have used (\ref{waper}).

Now, above result is certainly not true and it is easy to find where
we cheated. Splitting of configuration space (\ref{splcp}) imposes
specific boundary conditions 
\begin{displaymath}
\psi(2\pi,\sigma)=\psi(0,-\sigma),\quad
\psi'(2\pi,\sigma)=\psi'(0,-\sigma)
\end{displaymath}
which are obviously incompatible with the definition of
$``\sigma_z"$ (\ref{sigz}). But, if we restrict ourself to low
energy perturbation theory around classical minimum, then boundary
conditions became irrelevant and we can consider our \emph{purely
bosonic} system to behave as $N=2$ SUSY theory (\ref{h2pi}).

The same thing can be seen looking at the ``smoothness'' properties
of $``\sigma_z"$. From its definition we can see that when it acts
on eigenvectors of $\hat{q}$, its eigenvalue jumps from $1$ to
$-1$ when $q$ passes $2\pi$. From that we can conclude that
$``\sigma_z"$, and so fermion number $F$ also, can be defined only
locally around classical minima.

\subsection{Summary}

If we want a good classical physics limit continuous under sign
change of $\hbar$, i.e., under $\hbar\to-\hbar$, then if a wave
packet jumps from $x\to f(x)$ we must interprete that $x$ and $f(x)$
are (after all, we pretend it) the same point. Otherwise the
classical position is not smoothly going with $\hbar$.

To pretend such identification of points without making the
(classical) mechanical properties of the particle jump, the map $f$
must be an isometry and the potential must be invariant:
\begin{displaymath}
V(f(q))=V(q)
\end{displaymath}
or say,
\begin{displaymath}
V^{(0)}(q)=\left(W'(q)\right)^2=\left(W'(f(q))\right)^2=V^{(0)}(f(q))
\end{displaymath}
Let us emphasize main point again. As you vary the parameters -- say
$\hbar$ -- so that near a minimum (or for that matter near wave
function $\psi_g$) its $\ln\psi(q)$ change sign, then the wave
packet will jump somewhere far away, or become non-normalizable (in
non-compact case). This jump must by ``identification'' be pretended
not to occur.

To live up to the requirements of smoothness under $\hbar$
continuing to $-\hbar$, the place where a wave packet jumps under
$\hbar\to-\hbar$ must (classically at least) behave like the place it
jumped from. It follows that $q\to f(q)$ decribing the jumping of
narrow wave packages must be a (classical) symmetry transformation
of the configuration space.

As a simple example we considered the $4\pi$-periodic pure quantum
mechanical system with no fermionic degrees of freedom. We have
shown that it is equivalent to a $2\pi$-classically-periodic
system with a fermionic degree of freedom which:
\begin{itemize}
\item has exact SUSY with Hamiltonian given by
\begin{displaymath}
H=Q^2+\mbox{constant}
\end{displaymath}
where SUSY generator $Q$ is
\begin{displaymath}
Q=Q^\dagger=\left(\hat{p}+iW'(\hat{q})\right)e^{\frac{i}{\hbar}2\pi
\hat{p}}=\left(\hat{p}+iW'(\hat{q})\right)``\sigma_x"
\end{displaymath}
\item has -- but only locally, or to perturbative approximation to
all orders -- a conserved fermion number with
\begin{displaymath}
(-1)^F=``\sigma_z"
\end{displaymath}
where $``\sigma_z"$ is distinguishing points in configuration space
wich are classically indistinguishable, i.e., $q$ from $q+2\pi$.
\end{itemize}

\section{Conclusion}

For the purpose of construction of realistic models it would
be desirable to construct nonsupersymmetric theories having certain
simple properties which are usually consequences of supersymmetry,
e.g., vanishing of the cosmological constant.
It is therefore that we investigate in this paper consequences of
certain simple assumptions on quantum mechanical models. We assume
smoothness of ground state properties in Planck constant and
vanishing of the ground state energy (at least perturbatively). In
fact, these two properties are related because vacuum fluctuations
produce kink type singularities.

We start from a classical bosonic theory. The resulting Hamiltonian
consists of a classical bosonic part with a potential of the form
$(W')^2$, and an additional $\hbar$ term of the form $-\hbar W''$.
The function $W$ is called superpotential. The absolute value of the
last term is exactly equal to vacuum fluctuation term but the
term itself changes sign from one classical vacuum to the next one.
Thus in the case of many degenerate vacua due to positive definiteness
of vacuum fluctuations the desired property will be fulfilled only in
half of the minima. It is thus impossible to fulfill our
assumptions with the pure bosonic theory. Complete cancelation in
all minima can be however obtained by doubling the Hilbert space of
states and adding the term $\hbar W''\sigma_3$. The result is SUSY
QM. Such a procedure can be generalised to quantum mechanics of $n$
bosonic degrees of freedom (Section 3.3). The requirement of
subtraction of quantum fluctuations for all critical points of the
superpotential leads to restrictions (see Eq. (\ref{Dcond}) in Section
3.3) which are solved by SUSY Hamiltonian. In the case of 2
dimensional quantum mechanics it is shown to be the only solution
(Section 3.3.1). This analysis was done for generic superpotential
with generic critical points. It is interesting to consider a
superpotential with some constraints on critical points. In particular
a superpotential is taken which has all critical points with index 1
and traceless Hessian (see Eq. (\ref{wzh}) in Section 3.4). In that
case we obtain by the subtraction procedure the Hamiltonian
(\ref{wzhc}). This Hamiltonian is related to the Wess-Zumino $N=4$
SUSY QM (\ref{wzqm}). In fact they coincide in the bosonic sector
(fermion number 0 and 2). The WZ model has in addition a fermionic
sector with nonzero vacuum energy. Here the SUSY terms vanish.

The previous analysis was first performed in first order in $\hbar$.
One would naturally expect it to be true also in higher orders. In
Appendix A was performed the second order analysis for the
particular case of one dimensional bosonic quantum mechanics. It was
again shown that vacuum energy cancellation leads to bosonic part of
SUSY Hamiltonian.

In the Section 4 we have taken a slightly different point of view.
We have assumed certain smoothness assumptions (in fact we 
assume a
classical limit to hold even letting $\hbar$ go to be -- small and --
negative) and have then shown under the rather strong consequences
restricting the properties of the purely bosonic compact QM considered
to be perturbatively equivalent to a SUSY quantum system. These strong
consequences imposed include several -- at least two -- minima in the
potential $V$, and a discrete symmetry reflecting in some way the
classical configuration space. Then the fermionic degree of freedom is
identified with the label(s) separating the components of the
configuration space into which we divide it to present at the end the
classical configuration space as one of these components, the other
one being an identified copy using the symmetry as identification
(made up to get rid of jumping of the wave packet under
$\hbar\to-\hbar$).

Finally, previous analysis would suggest that certain simple
assumptions like smoothness of ground state properties in $\hbar$
or vanishing of ground state energy would require supersymmetry.
That would mean that it is very difficult to avoid SUSY and if that
is necessary because of phenomenological reasons one has to abandon
also previously mentioned properties. It is also important to stress
that a particular consequence of previous statements is that bosons
without fermions cannot satisfy above requirements, at least in
cases considered.

It would be interesting to pursue further investigations in quantum
mechanics and field theory to see how general these
conclusions could be or could they be avoided in some
circumstances.

\begin{ack}
Two of us (S.\ P. and P.\ P.) would like to acknowledge the kind
hospitality of CERN Theory Division where final work on this paper
was done. We would like also to acknowledge the financial support
under the contract No. 119222 of Ministery of Science and
Technology of Republic of Croatia. H.\ B.\ N. thanks the EU
commision for grants SCI-0430-C (TSTS), CHRX-CT-94-0621,
INTAS-RFBR-95-0567 and INTAS 93-3316 (ext). We would also like to
thank Don Bennett for many discusions.
\end{ack}

\appendix

\section{Appendix: Perturbative analysis in 1D QM}

We start from one-dimensional Hamiltonian
\begin{displaymath}
H=\frac{p^2}{2}+V(x)
\end{displaymath}
It is well-known that if the ground state energy is discrete and
\emph{exactly} equal to zero\footnote{If this is true for every
value of $\hbar$ it follows that $V(x)$ explicitely depends on
$\hbar$.}, the potential $V(x)$ can always be
written in the form of the Riccati equation
\begin{equation}\label{ricc}
V(x)=\frac{1}{2}\left(W(x)\right)-\frac{\hbar}{2}W'(x)
\end{equation}
where $W(x)$ is determined from the ground state wave function
$\psi_v(x)$
\begin{displaymath}
W(x)=-\hbar\frac{d}{dx}\ln\psi_v(x)
\end{displaymath}

Now we can ask ourself if that is also true if we instead demand the
weaker condition that ground state energy is equal to zero only
\emph{perturbatively} in $\hbar$. That means the following.

We start from the classical (i.e., independent of $\hbar$) potential
$V_0(x)$ which is positive, so we can write it as $V_0=W_0^2/2$. We
assume that the classical vacuum energy is zero, so there is (at
least one) point $x_0$ in which $V_0(x_0)=0$ (from positivity
obviously follows $V_0'(x_0)=0$). We suppose that $x_0$ is a
nondegenerate critical point of $V_0$, i.e., $V_0''(x_0)\ne0$. From
$V''=(W_0')^2+W_0W_0''$ and $W_0(x_0)=0$ follows that $W_0'(x_0)\ne0$.
For notational simplicity we translate $x$ so that $x_0=0$.

After quantization the ground state energy obtains a quantum
correction which for such potential is strictly larger than zero. Now
we add $\hbar$ dependent term and take $V$ equal to (after expanding
in $\hbar$)
\begin{equation}\label{vhex}
V(x)=\sum_{n=0}^\infty\hbar^nV_n(x)
\end{equation}
Our goal is to find out conditions on $V_n$ which follow from
requirement that vacuum energy vanishes in every order of perturbation
expansion in $\hbar$. We shall make calculation to second order
and show that to this order obtained conditions on $V_n$ are exactly
those which follow from condition that $V$ can be written in the
form of the Riccati equation (\ref{ricc}) (where $W$ generaly depends
on $\hbar$).

To do perturbative expansion of vacuum energy in $\hbar$ we must
Taylor-expand potential $V(x)$ around classical vacuum $x=0$. For
this we need following expansions:
\begin{eqnarray*}
W_0(x)&=&\sum_{k=1}^\infty w^{(0)}_k x^k \\
V_n(x)&=&\sum_{k=0}^\infty v^{(n)}_k x^k
\end{eqnarray*}
Using ordinary perturbation theory and collecting terms wich are of
the same order in $\hbar$, we obtain expansion for vacuum energy
\begin{displaymath}
E_v=\sum_{n=1}^\infty\hbar^n\Delta_n
\end{displaymath}
Requiring $\Delta_n=0$, $\forall n$ leads to constraints on Taylor
coefficients $v^{(n)}_k$.

Now we present results for first two terms.

\subsection{1st order}

\begin{displaymath}
\Delta_1=\frac{1}{2}+\frac{v^{(1)}_0}{|w^{(0)}_1|}
\end{displaymath}
From $\Delta_1=0$ follows the condition
\begin{equation}\label{1stord}
v^{(1)}_0=-\frac{|w^{(0)}_1|}{2}
\end{equation}

\subsection{2nd order}

\begin{displaymath} \!\!\!\!\!\!\!\!\!\!\!
2\Delta_2=\frac{3}{2}\frac{\eta w^{(0)}_3}{\left(w^{(0)}_1\right)^2}
+\frac{v^{(1)}_2}{\left(w^{(0)}_1\right)^2}-\frac{\eta}{\left(
w^{(0)}_1\right)^3}\left(v^{(1)}_1+\eta w^{(0)}_2\right)\left(
v^{(1)}_1+2\eta w^{(0)}_2\right)+\frac{2v^{(2)}_0}{|w^{(0)}_1|}
\end{displaymath}
where $\eta=w^{(0)}_1/|w^{(0)}_1|$.
From $\Delta_2=0$ now follows the condition
\begin{equation}\label{2ndord}
v^{(1)}_2=-2\eta w^{(0)}_1v^{(2)}_0+\frac{\eta}{w^{(0)}_1}
\left(v^{(1)}_1+\eta w^{(0)}_2\right)\left(v^{(1)}_1+2\eta
w^{(0)}_2\right)-\frac{3}{2}\eta w^{(0)}_3
\end{equation}

\subsection{``Riccati conditions''}

Now we want to see what are the conditions on $V(x)$ if it can be
written in ``Riccati form'' (\ref{ricc}). For that we must first
expand ``superpotential'' $W(x)$ in $\hbar$
\begin{displaymath}
W(x)=\sum_{n=0}^\infty\hbar^n W_n(x)
\end{displaymath}
Riccati Eq. (\ref{ricc}) now takes the form (\ref{vhex}) where
$V_n(x)$ are given by
\begin{equation}\label{vwn}
V_n(x)=\frac{1}{2}\sum_{m=0}^nW_m(x)W_{n-m}(x)-\frac{1}{2}W_{n-1}'(x)
\end{equation}
Our goal is to obtain relations between Taylor coefficients
$v^{(n)}_k$ of $V_n(x)$. When we put expansion
\begin{displaymath}
W_n(x)=\sum_{k=0}^\infty w^{(n)}_k x^k
\end{displaymath}
into (\ref{vwn}) we obtain
\begin{displaymath}
v^{(n)}_k=\frac{1}{2}\sum_{m=0}^n\sum_{j=0}^k w^{(m)}_j
w^{(n-m)}_{k-j}-\frac{k+1}{2}w^{(n-1)}_{k+1}
\end{displaymath}
We give first few terms explicitely
\begin{eqnarray}
v^{(1)}_0&=&-\frac{w^{(0)}_1}{2} \label{v10} \\
v^{(1)}_1&=&w^{(0)}_1w^{(1)}_0-w^{(0)}_2 \label{v11} \\
v^{(1)}_2&=&w^{(1)}_0w^{(0)}_2+w^{(0)}_1w^{(1)}_1-\frac{3}{2}
w^{(0)}_3 \label{v12} \\
v^{(2)}_0&=&\frac{1}{2}\left(w^{(1)}_0\right)^2-\frac{w^{(1)}_1}{2}
\label{v20} \\
v^{(2)}_1&=&w^{(1)}_0w^{(1)}_1+w^{(0)}_1w^{(2)}_0-w^{(1)}_2
\label{v21}
\end{eqnarray}
If we take $W_0(x)$ ($w^{(0)}_k$) as given and fixed, we see that
(\ref{v10}) is a condition on potential $V(x)$. From 
(\ref{v11}--\ref{v20}) we can easily obtain second condition
\begin{equation}\label{v12f}
v^{(1)}_2=-2w^{(0)}_1v^{(2)}_0+\frac{1}{w^{(0)}_1}
\left(v^{(1)}_1+w^{(0)}_2\right)\left(v^{(1)}_1+2w^{(0)}_2\right)
-\frac{3}{2}w^{(0)}_3
\end{equation}
Because (\ref{v11}--\ref{v20}) are three relations including two
free parameters ($w^{(1)}_0$ and $w^{(1)}_1$), (\ref{v12f}) is the
only relation between involved parameters.

Now, if $w^{(0)}_1>0$ ($\eta=1$) we can see that (\ref{v10}) is
equal to (\ref{1stord}), and (\ref{v12f}) is equal to (\ref{2ndord}).
That means that conditions which follow from requirement of
perturbative (in $\hbar$) vanishing of vacuum energy are equivalent
with those imposed by condition that potential can be written in the
Riccati equation form (\ref{ricc}), at least in first two orders.

\end{document}